\documentclass[reprint,aps]{revtex4-2}

\usepackage[centertags]{amsmath}
\usepackage{graphicx}
\usepackage{amsfonts}
\usepackage{epstopdf}
\usepackage{color} 
\usepackage{cancel}
\usepackage{verbatim}
\usepackage{caption}
\usepackage{subcaption}
\usepackage{float}

\begin{document}

\title{Extreme wave excitation from localized phase-shift perturbations}

\author{Y. He$^{1}$} 
\email{yuchen.he@sydney.edu.au}
\author{A. Witt$^{2}$} 
\email{andy.witt@tuhh.de}
\author{S. Trillo$^{3}$}
\author{A. Chabchoub$^{4,5,1}$} 
\author{N. Hoffmann$^{2,6}$} 
\email{norbert.hoffmann@tuhh.de}

\affiliation{$^1$ Centre for Wind, Waves and Water, School of Civil Engineering, The University of Sydney, Sydney, NSW 2006, Australia} 
\affiliation{$^2$ Dynamics Group, Hamburg University of Technology, Hamburg, Germany}
\affiliation{$^3$ Department of Engineering, University of Ferrara, via Saragat 1, 44122 Ferrara, Italy}
\affiliation{$4$ Hakubi Center for Advanced Research, Kyoto University, Yoshida-Honmachi, Kyoto 606-8501, Japan} 
\affiliation{$^5$ Disaster Prevention Research Institute, Kyoto University, Kyoto 611-0011, Japan} 
\affiliation{$^6$ Department of Mechanical Engineering, Imperial College London, London, United Kingdom}

\begin{abstract}
The modulation instability is a focusing mechanism responsible for the formation of strong wave localizations not only on the water surface, but also in a variety of nonlinear dispersive media. Such dynamics is initiated from the injection of side-bands, which translate into an amplitude modulation of the wave field. The nonlinear stage of unstable wave evolution can be described by exact solutions of the nonlinear Schrödinger equation (NLSE). In that case, the amplitude modulation of such coherent extreme wave structures is connected to a particular phase-shift seed in the carrier wave. In this letter, we show that phase-shifts localization applied to the background, excluding any amplitude modulation excitation, can indeed trigger extreme events. 
Such rogue waves can be for instance generated by considering the parametrization of fundamental breathers and thus, by seeding only the local phase-shift information to the regular carrier wave. 
Our wave tank experiments show an excellent agreement with the expected NLSE hydrodynamics and confirm that even though delayed in their evolution, breather-type extreme waves can be generated from a purely regular wave train. Such novel focusing mechanism awaits experimental confirmation in other nonlinear media, such optics, plasma, and Bose-Einstein condensates. 
\end{abstract}

\maketitle  
The formation of extreme wave events, so-called rogue waves, can be understood and modelled by the nonlinear Schrödinger equation (NLSE) or its higher-order forms \cite{akhmediev1997solitons,osborne2010nonlinear,onorato2013rogue,dudley2014instabilities}. One way to control such rogue waves in a laboratory environment is the use of fundamental breather solutions in form of time-periodic, doubly-localized, and space-periodic forms. These are also known as Akhmediev (AB) \cite{akhmediev1985generation}, Peregrine (PB) \cite{peregrine1983water}, and Kuznetsov breather (KB) \cite{kuznetsov1977solitons}, respectively. Such solutions are based on long wave perturbation of a regular wave train with a precise dedicated phase-shift of the carrier at a specific location and instant. 
While, modulational instability (MI) is usually seen as a mechanism that leads, through the amplification of sidebands, to a strong {\em amplitude modulation}, these phase-shifts always accompany such MI-induced amplitude reshaping.
Indeed, the initial temporal phase profile, or equivalently, the input phase relationship in Fourier space (i.e. the relative phase between the MI side-bands and the central carrier) have deep impact on the type of amplitude modulation
that develops upon propagation \cite{trillo1991dynamics,armaroli2020stabilization}, as recently observed experimentally \cite{Kimmoun2016,Mussot2018,Pierangeli2018}.

The aim of this paper is to show that the initial phase-shift profile contains the leading information that allows the breathers to undergo the peak growth and focusing process even when the input amplitude modulation is suppressed.
A previous experimental study has revealed that locally the same type of maximal breather compression can be achieved by just starting from the amplitude modulation only and ignoring the phase-shift in the boundary condition \cite{chabchoub2020phase}. 
In this experimental study, we provide evidence for the opposite phenomenon, namely that extreme wave events can be triggered and excited by a localized phase-shift perturbation (PSP) in a regular (i.e., constant amplitude) wave train. 

In our study, we use specific phase-shifts proper to the three fundamental breathers, as mentioned above, and show that the extreme localization achieved in the wave flume resembles those of pure breathers, though is spatially delayed with respect to the exact solution. In fact, the smaller the phase-shift applied to the carrier the better the agreement with the respective pure breather solution. All observations are in reasonable agreement with the NLSE prediction despite the very strong wave focusing reached. 

The NLSE is the simplest framework to describe the wave propagation of nonlinear dispersive waves \cite{ablowitz2011nonlinear}. For the deep-water wave problem the slowly-varying wave envelope $\Psi$ around the wave number $k$ satisfies \cite{zakharov1968stability,osborne2010nonlinear}
\begin{eqnarray} 
i\left(\Psi_x+\dfrac{2k}{\omega}\Psi_t\right)-\dfrac{k^2}{\omega}\Psi_{tt}-k^3\left|\Psi\right|^2\Psi=0,
\label{NLSE}
\end{eqnarray}
where $\omega=\sqrt{gk}$ is the dispersion relation, $g$ being the gravitational acceleration. 
\onecolumngrid
\begin{figure*}
\centering
\includegraphics[width=0.325\textwidth]{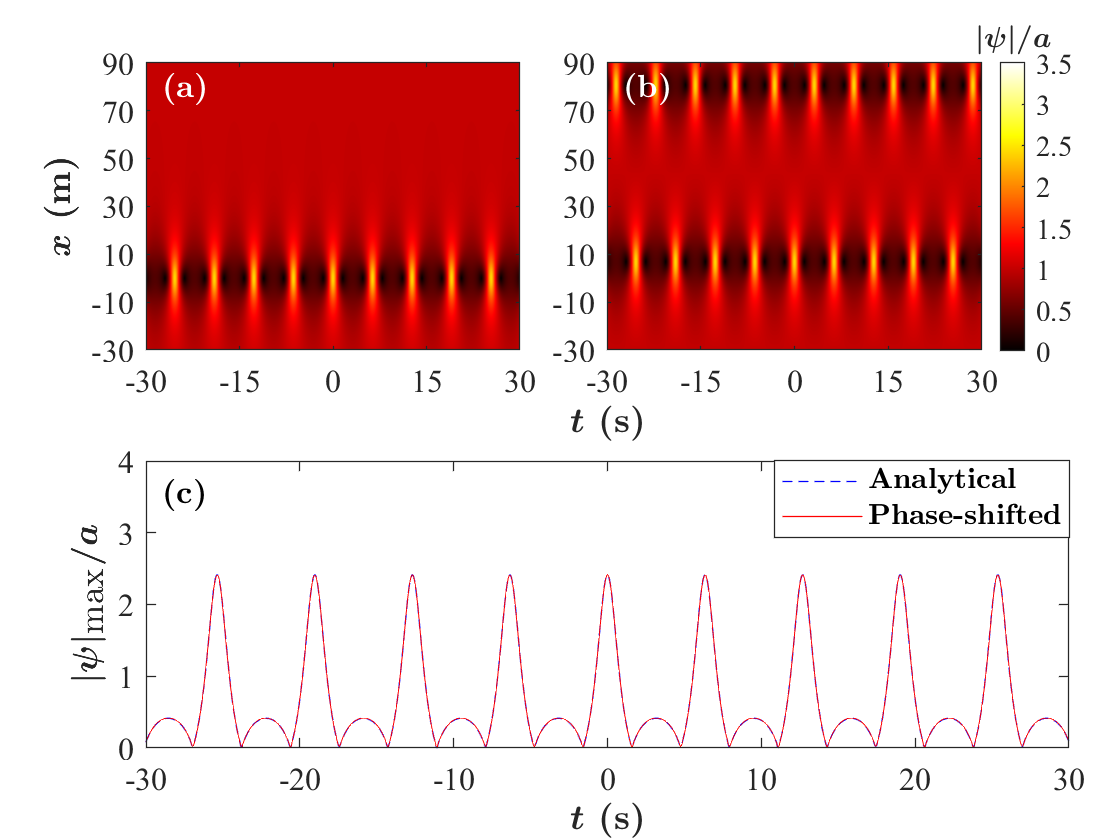}\includegraphics[width=0.325\textwidth]{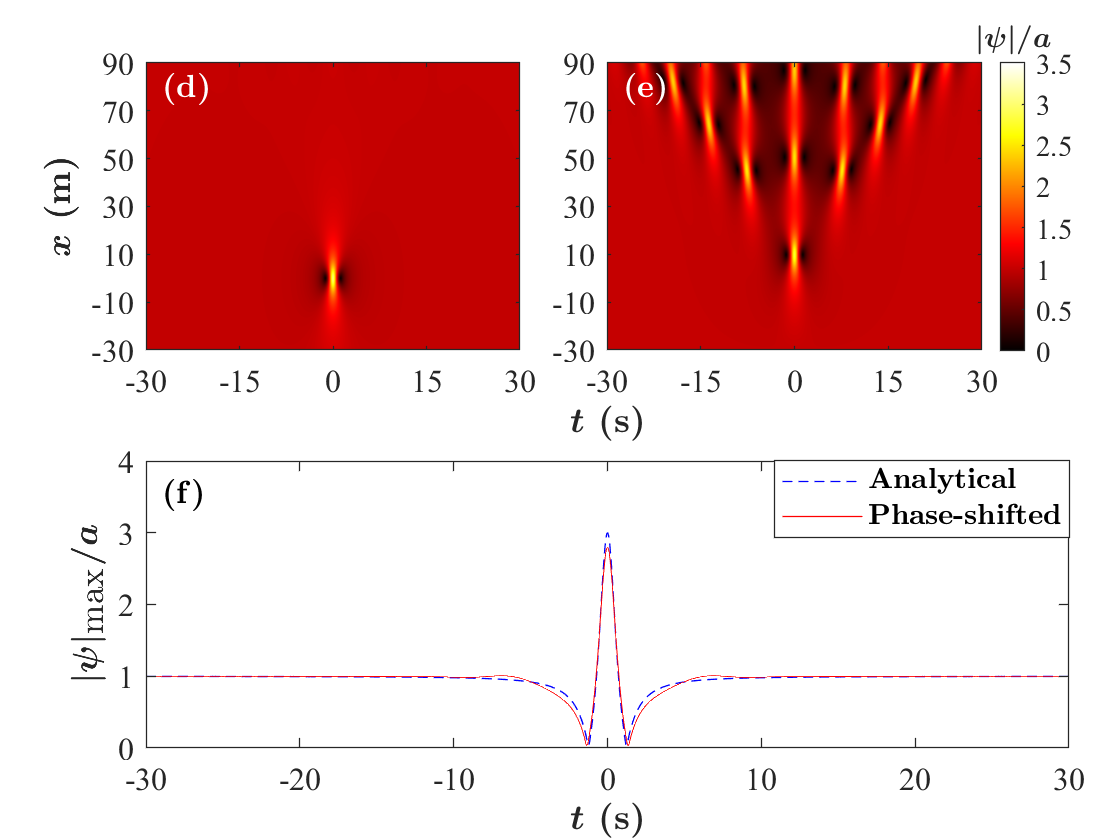}\includegraphics[width=0.325\textwidth]{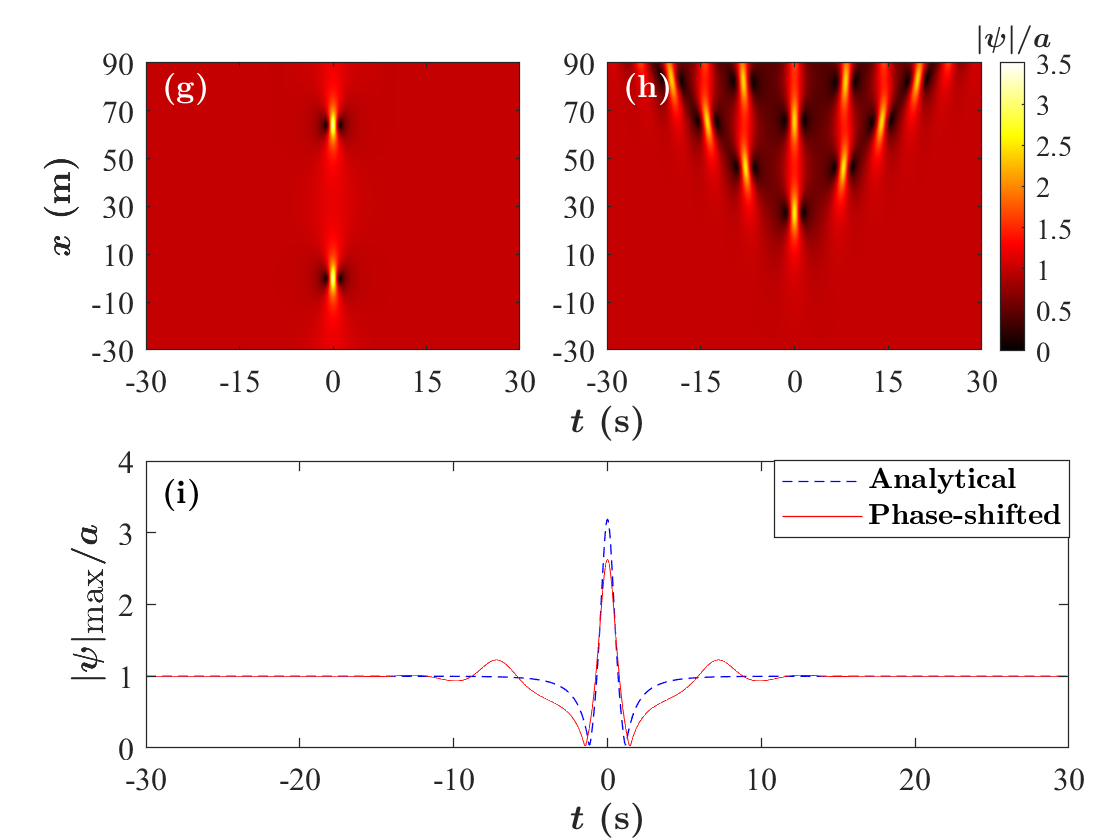}\onecolumngrid
\caption{
(a,b,c) Akhmediev breather, contrasting spatio-temporal evolutions arising from (a) the full analytical Solution and (b) the PSP; (c) temporal profile of exact solution compared to that generated from the PSP at the location of maximal compression.
(d,e,f) Peregrine breather, contrasting spatio-temporal evolutions arising from (d) the full solution and (e) the PSP; (f) comparison of transverse profiles at maximal compression.
(g,h,i) Kuznetsov breather, contrasting spatio-temporal evolutions arising from (g) the full solution and (h) the PSP; (i) comparison of transverse profiles at maximal compression.
Here the adopted carrier wave steepness $ak=0.1$ and carrier amplitude $a=0.01$ m. All waves are excited at $x=-30$, i.e. 30 meters ahead of the first focus point ($x=0$) of the exact solution. 
}
\label{fig1}
\end{figure*}
\twocolumngrid
As a result of integrability a variety of steady and pulsating solution have been derived \cite{shabat1972exact,akhmediev2021waves}. We will not recall the parametrization of three fundamental breathers on a finite-amplitude carrier as these have been discussed in many past publications \cite{dudley2010modulation,kibler2010peregrine,kibler2012observation,chabchoub2016hydrodynamic}. The family of ABs describe the nonlinear stage of MI and are periodic in time when considering the framework (\ref{NLSE}), the PB is the limit case of zero modulation frequency, i.e. infinite modulation period, and the KBs the space periodic solution, which starts its evolution from a solitonic perturbation on the carrier. As such, the KB never converges to a uniform amplitude. All these solutions have been observed in a variety of nonlinear dispersive media and, for more details, we refer the reader to  \cite{dudley2010modulation,kibler2010peregrine,chabchoub2011rogue,kibler2012observation,chabchoub2016hydrodynamic}. In water waves the boundary conditions at beginning of the flume consist of time-series of surface displacement. The latter are driving the wave maker while its mechanical motion is linearly proportional to the surface elevation signal as determined by theory. To drive the wavemaker and to observe NLSE solitons or breathers, it is sufficient to use the expression of surface elevation defined to first-order of approximation \cite{chabchoub2011rogue}
\begin{eqnarray} 
\eta(x,t)=\textnormal{Re}\left(\Psi\left(x,t\right)\exp\left[i\left(kx-\omega t\right)\right]\right).
\label{se}
\end{eqnarray}
As mentioned in \cite{chabchoub2020phase}, each of these exact solutions $\Psi(x,t)$ can be written as amplitude modulation function $A(x,t)$ and a respective phase-shift $\phi(x,t)$, both evolving in time and space. Mathematically speaking, $\Psi(x,t)=A(x,t)\exp[i\phi(x,t)]$. Also in the latter work, it has been shown that wave breather-like focusing can develop when starting from boundary conditions at beginning of the flume, involving the amplitude modulation only without consideration of the breather-specific local phase-shift.
As next, we propose a novel extreme wave focusing mechanism, which originates in a regular and amplitude-modulation-free wave train, with a PSP, i.e. a localized phase-shift. While a similar mechanism was originally investigated theoretically in \cite{sheveleva2020temporal} for abrupt phase jumps, in this paper, we rather consider the PSP obtained by impressing the localized phase profile of the pure breather solution at a given distance from the focus point. Examples of evolution, obtained from the NLSE (1) are shown in Fig. \ref{fig1}(a,b,c) for the AB, in Fig. \ref{fig1}(d,e,f) for the PB, and in Fig. \ref{fig1}(g,h,i) for the KB. In all examples $x=0$ stands for the first focus point, and the PSP is excited at $x=-30$. As shown, in all three different breather cases considered, a strong wave envelope compression is expected to occur in the condensate from a PSP-triggering only. We emphasize that the smaller the from the solution defined amplitude modulation, and as such the related phase-shift of the exact solution for the smaller $x$ values, the better the agreement with the extreme waves starting from the phase-shift perturbation at the maximum focus point. Note that similarly to the mechanism studied in \cite{chabchoub2020phase}, the location of maxima deviates and is retarded compared to the motion of exact solution. Moreover, we can clearly notice that deviations are rather substantial when considering the KM case. The reason for this is that the solution never converges to the background and thus triggering the boundary conditions from a regular wave train with $\phi_{KM}(x^*,t)$ will more substantially deviate compared to the AB- or PB-type dynamics. In any event, the PSP-induced dynamics for both the PB and the KM cases show a post-focusing dynamics dominated by fissions into pairs of quasi-KM breathers with opposite velocities \cite{chabchoub2020phase}, whose observation, however, require tanks with length far exceeding that of the present experiment.

In the following we report on our experimental investigation. We adopt the same set-up as described in \cite{he2021phase}. Moreover, we have to be cautious about the fact that carrier parameters have to be chosen in such a way to allow for the observation of the first focus (compression) point within the tank length  for either the full exact breather or the corresponding respective PSP of the regular carrier wave.

\begin{figure}[ht]
\centering
\includegraphics[width=0.21\textwidth]{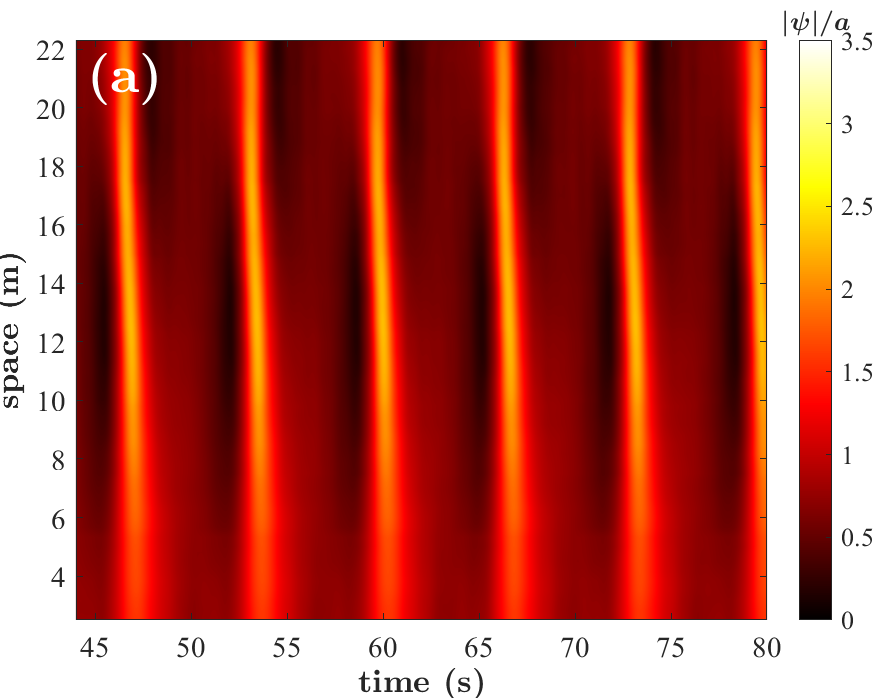}
\includegraphics[width=0.21\textwidth]{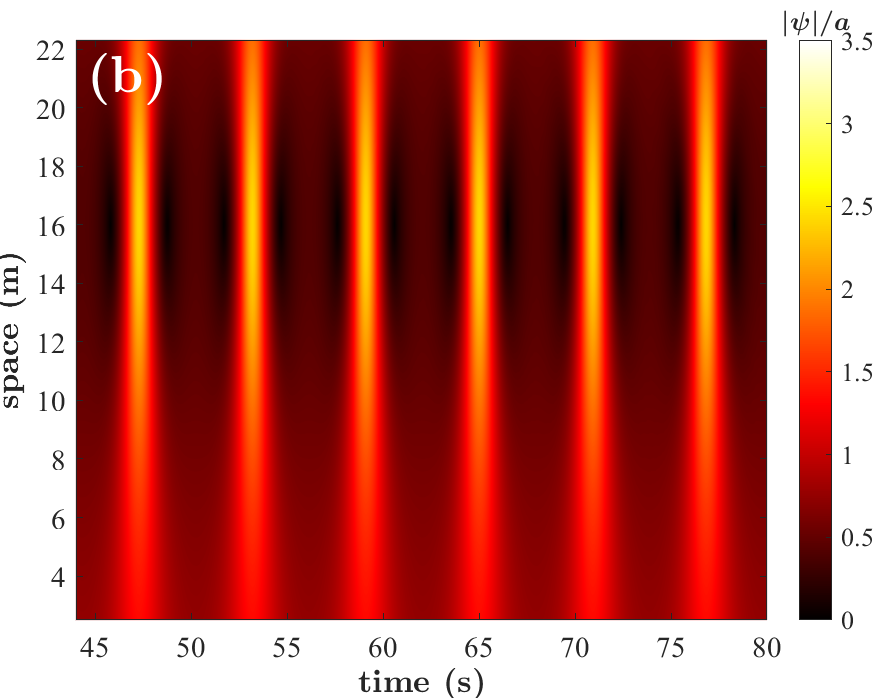}
\includegraphics[width=0.21\textwidth]{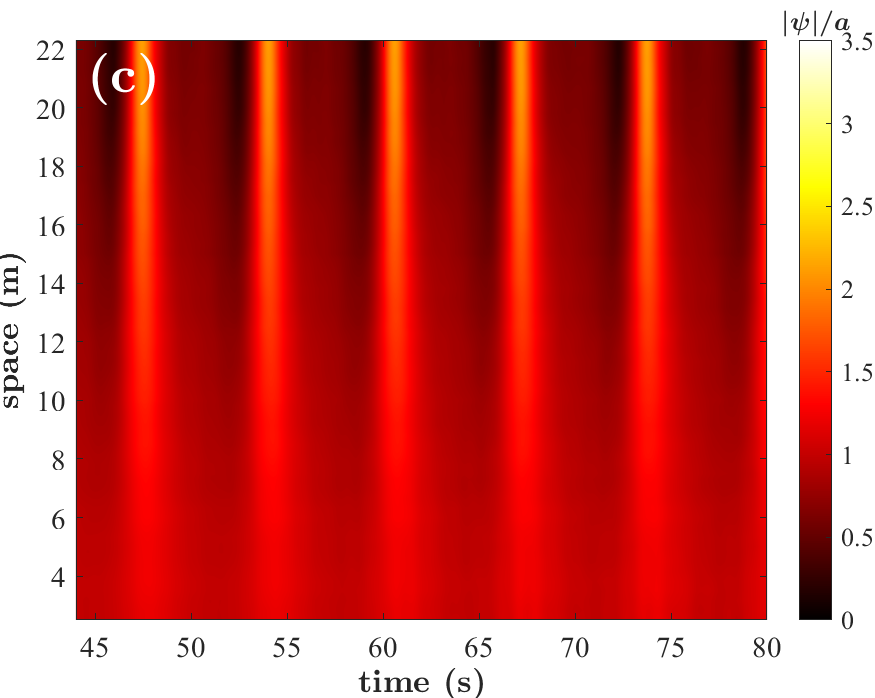}
\includegraphics[width=0.21\textwidth]{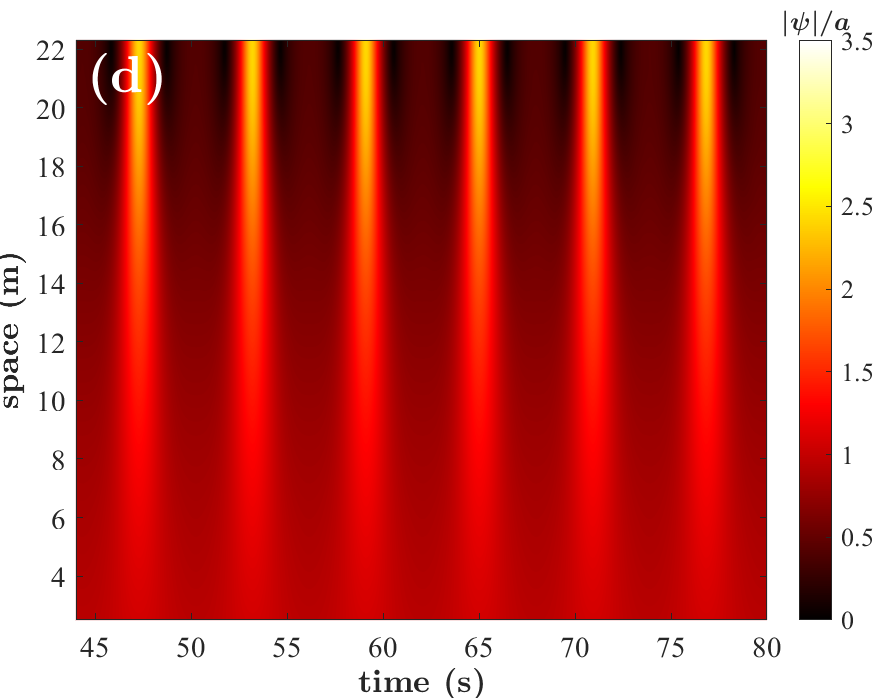}
\includegraphics[width=0.42\textwidth]{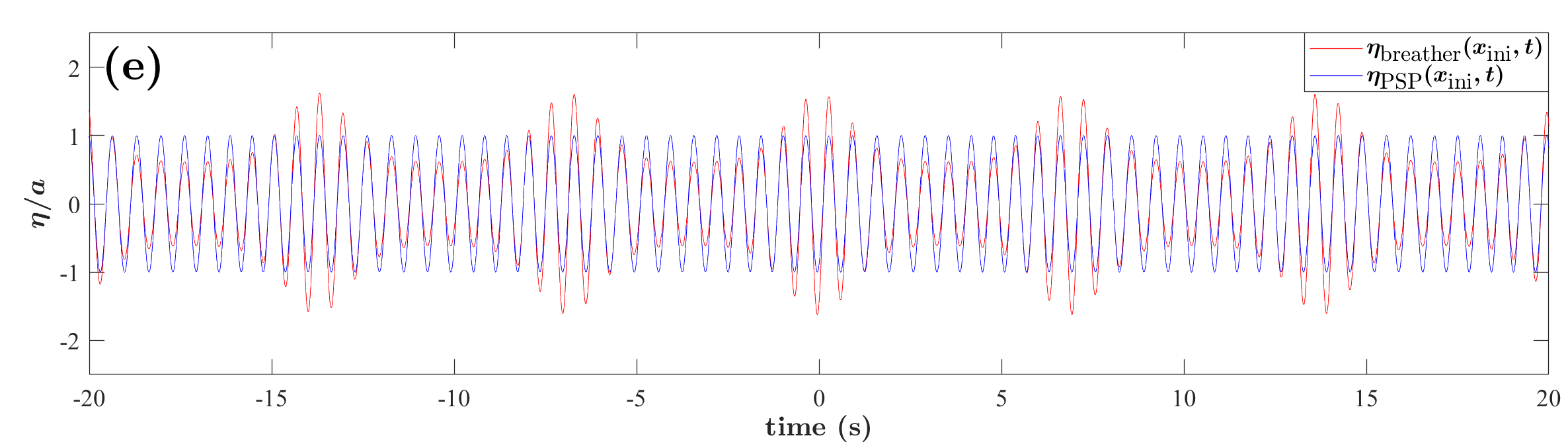}
\caption{AB evolution for $ak=0.12$ and $a=0.01$ with expected amplitude amplification of 2.41 (the case of maximal growth rate. (a) Propagation of wave envelope as measured in the flume. (b) NLSE simulations with same boundary conditions as in (a). (c) Experimental observation of the AB-PSP in the regular background case. (d) NLSE simulations with same boundary conditions as in (c). (e) Analytical and phase-shifted AB wave elevation boundary conditions at $x=-16$ m as implemented into the experiments and simulations.}
\label{fig2}
\end{figure}
As a starting point, our experimental investigation is initiated by investigating the AB dynamics triggered by both exact and PSP configurations as discussed above and illustrated in Fig. \ref{fig1}(a,b,c). The results are shown in Fig. \ref{fig2}. 
All collected measurements of the exact time-periodic breather show a reasonable agreement with NLSE theory. This applies for both the case when considering the pure AB solution of the NLSE and the regular wave train in which the AB-type phase-shift have been locally seeded. Remarkably, a gradual focusing clearly emerge from the phase-shift seeding in agreement with the numerical NLSE expectations. 






\begin{figure}[ht]
\centering
\includegraphics[width=0.21\textwidth]{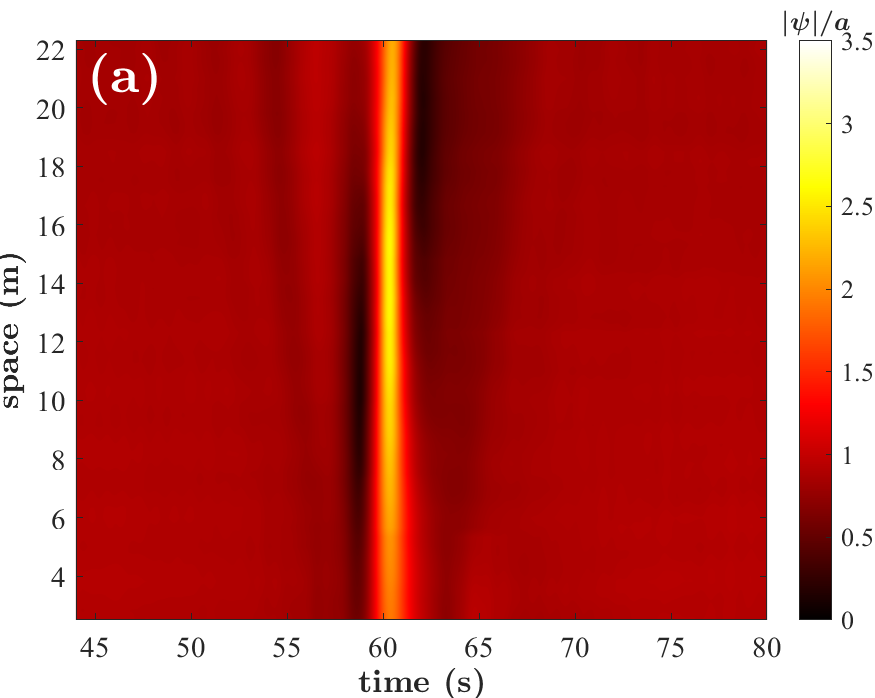}
\includegraphics[width=0.21\textwidth]{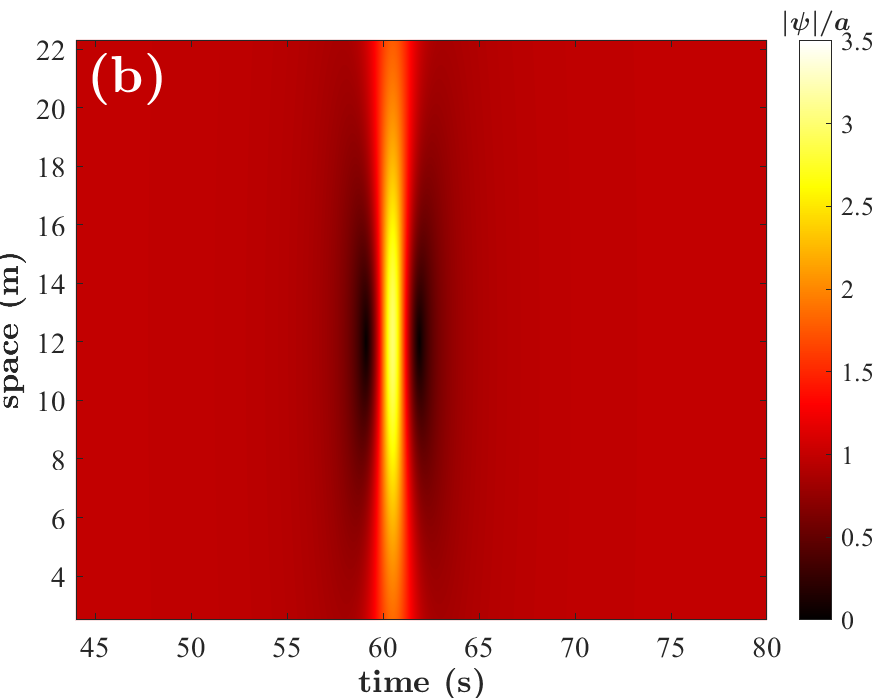}
\includegraphics[width=0.21\textwidth]{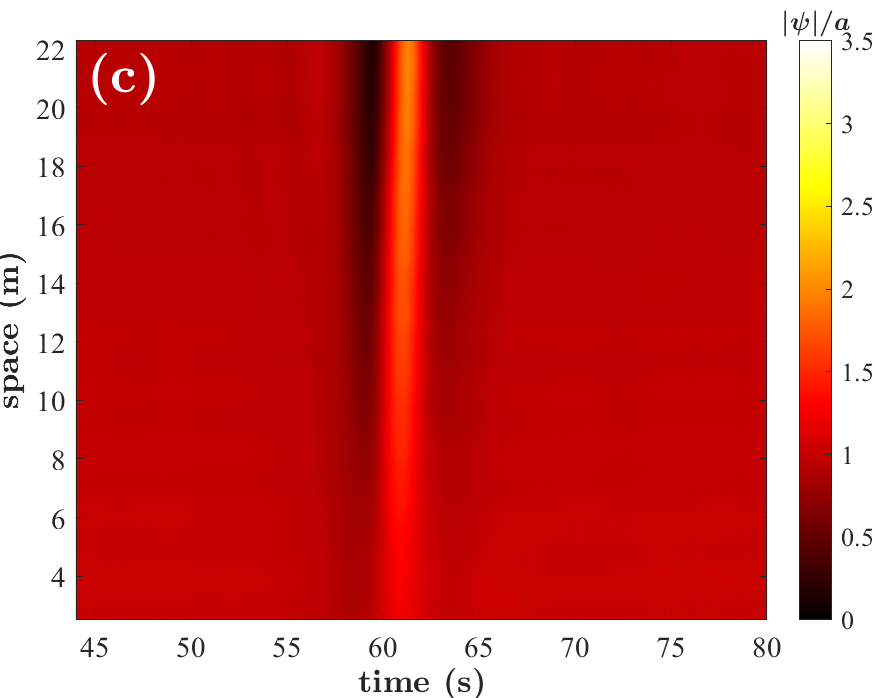}
\includegraphics[width=0.21\textwidth]{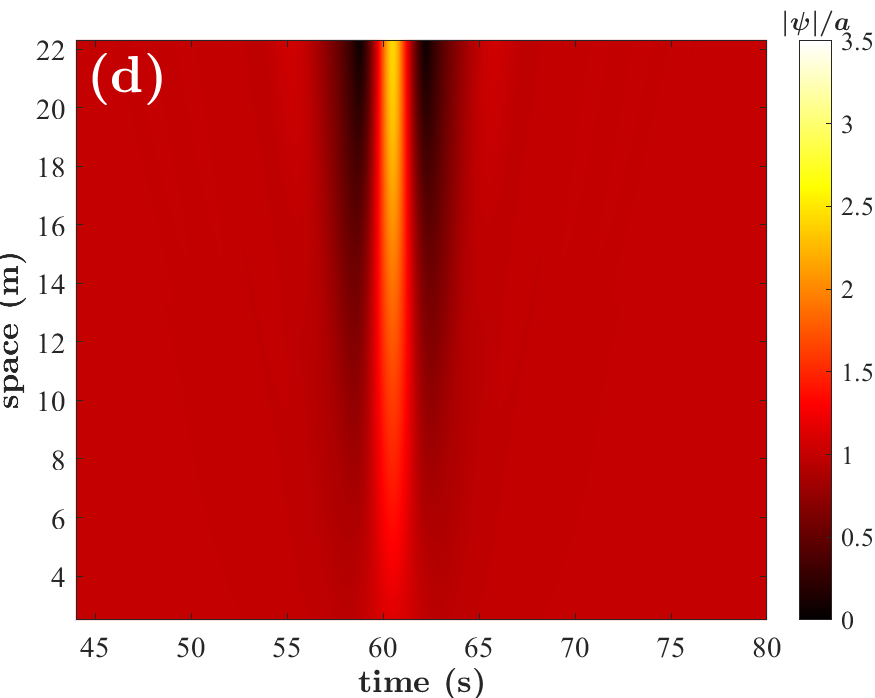}
\includegraphics[width=0.42\textwidth]{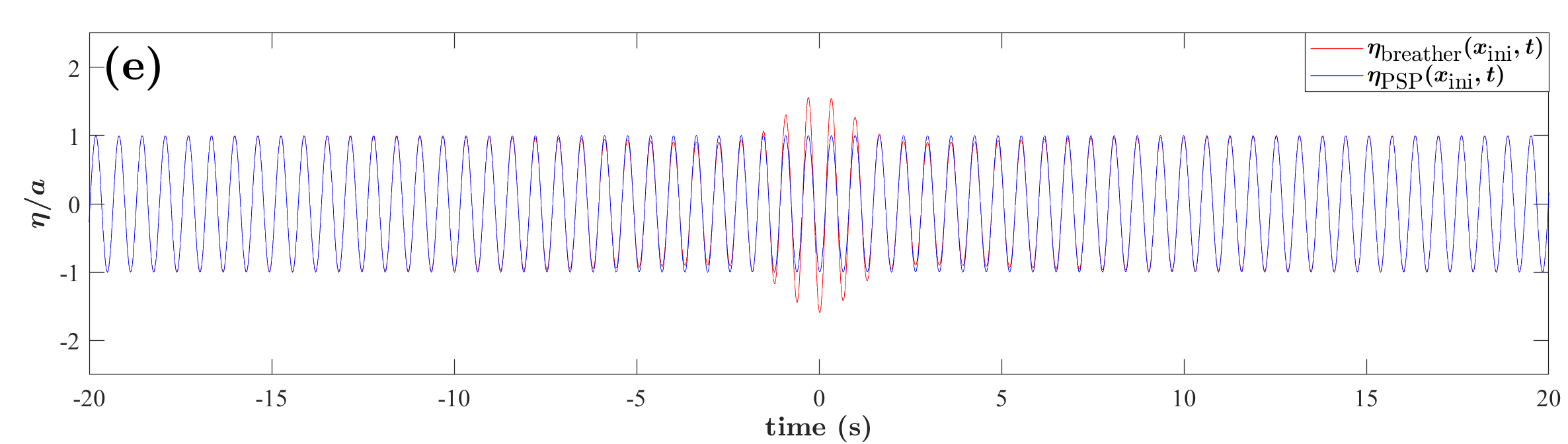}
\caption{PB evolution for $ak=0.1$ and $a=0.01$ with expected theoretical amplitude amplification of 3. (a) Propagation of wave envelope as measured in the flume. (b) NLSE simulations with same boundary conditions as in (a). (c) Experimental observation of the PB-PSS in the regular background case. (d) NLSE simulations with same boundary conditions as in (c). (e) Analytical and phase-shifted PB wave elevation boundary conditions at $x=-12$ m as implemented into the experiments and simulations.}
\label{fig3}
\end{figure}
\begin{figure}[h]
\centering
\includegraphics[width=0.21\textwidth]{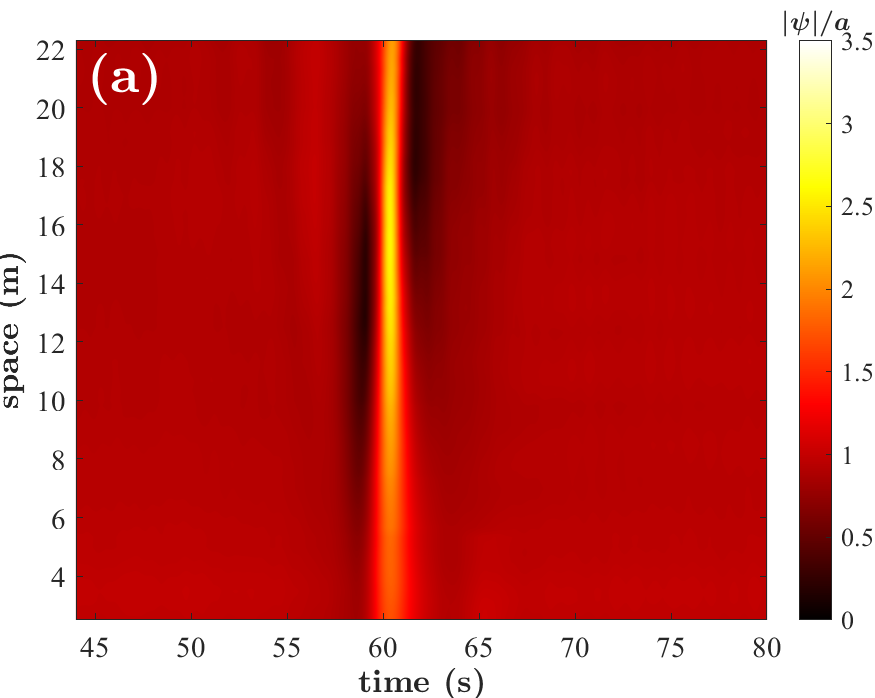}
\includegraphics[width=0.21\textwidth]{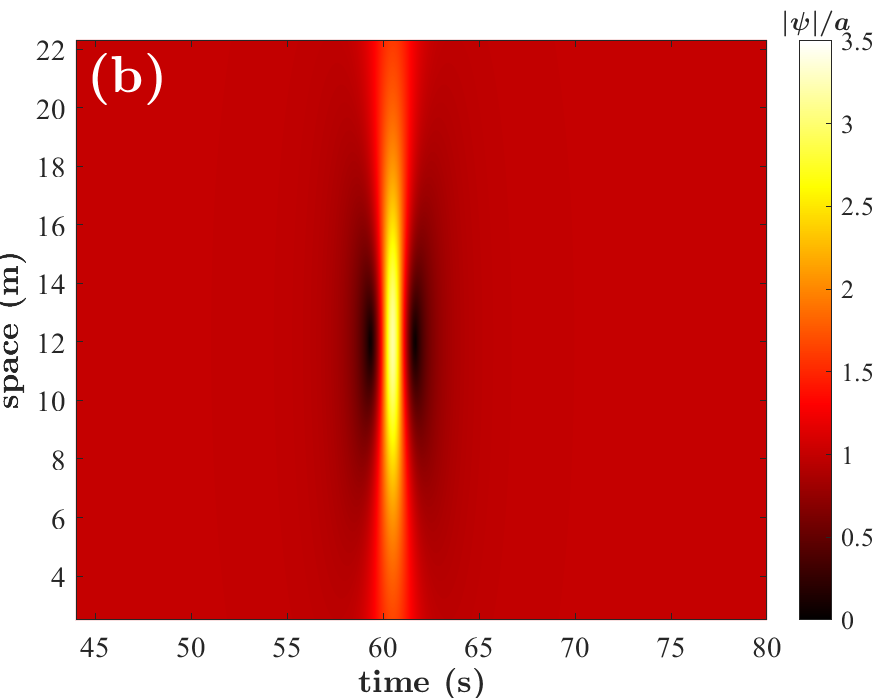}
\includegraphics[width=0.21\textwidth]{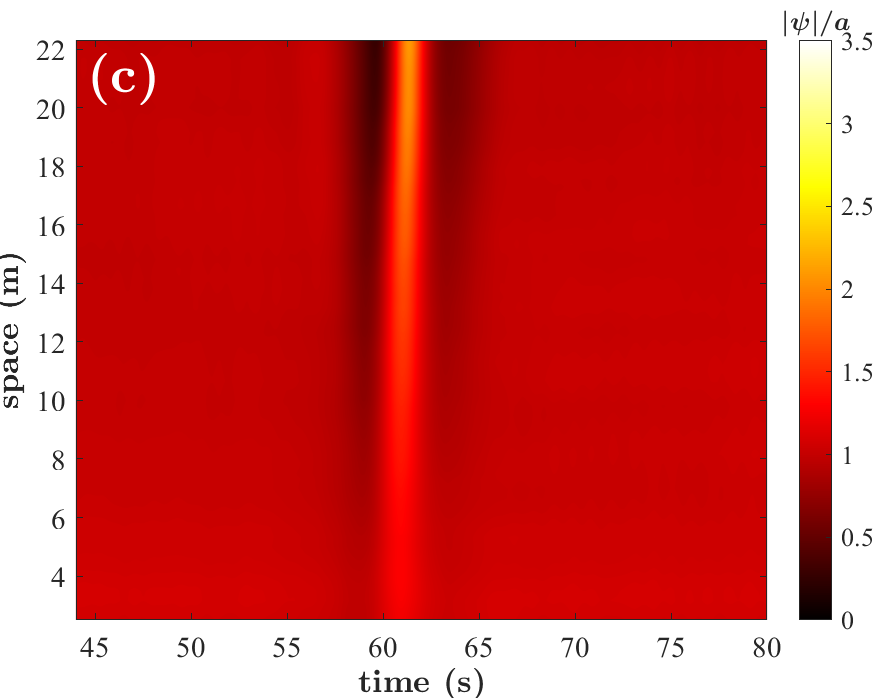}
\includegraphics[width=0.21\textwidth]{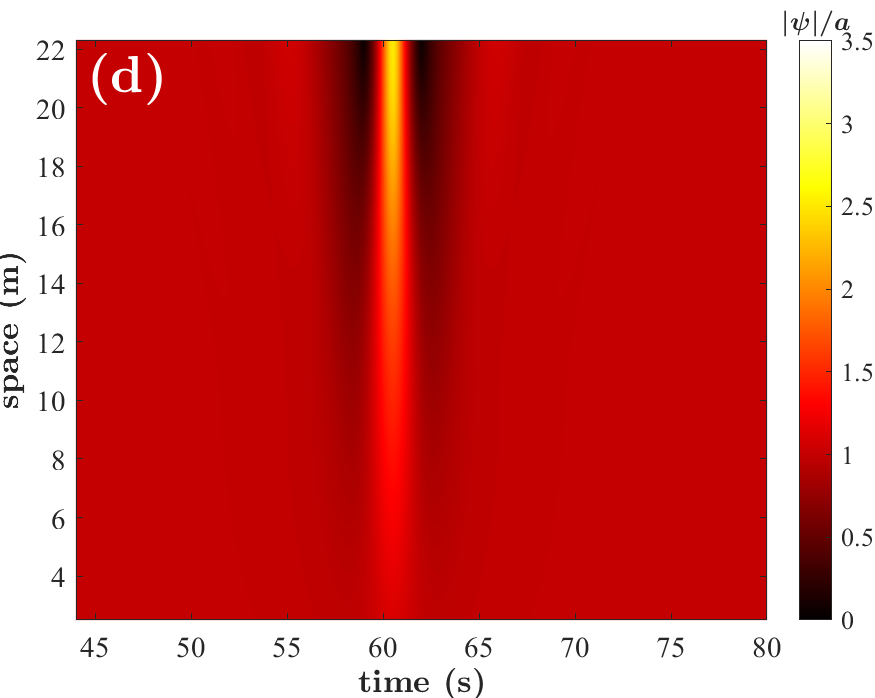}
\includegraphics[width=0.42\textwidth]{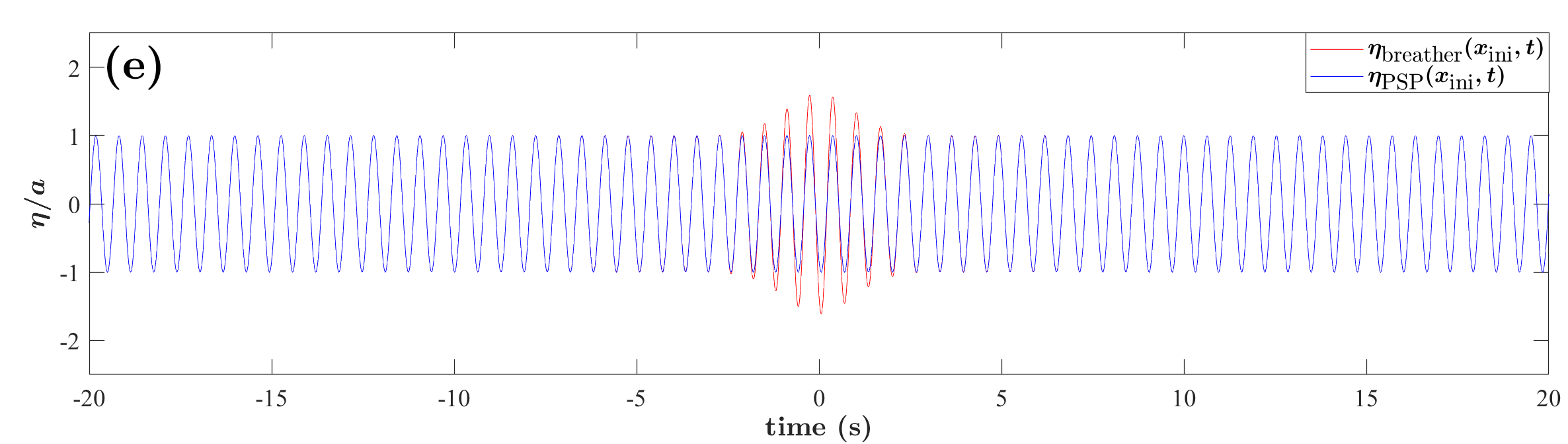}
\caption{KM evolution for $ak=0.1$ and $a=0.01$ with expected amplitude amplification of 3.3. (a) Propagation of wave envelope as measured in the flume. (b) NLSE simulations with same boundary conditions as in (a). (c) Experimental observation of the KM-PSS in the regular background case. (d) NLSE simulations with same boundary conditions as in (c). (e) Analytical and phase-shifted KM wave elevation boundary conditions at $x=-12$ m as implemented into the experiments and simulations.}
\label{fig4}
\end{figure}

The same excellent agreement in the wave focusing is also observed for case of PB, which is illustrated in Fig. \ref{fig3}.
In this case one particular feature of focusing from the PSP in the background of constant amplitude becomes clear. The retardation and delay in the focusing from the single unstable Peregrine-packet is noticeable. A similar feature has been observed when ignoring the breather-specific phase-shift and considering only the amplitude modulation \cite{chabchoub2020phase}. These observations also expose the limitations of our experimental set-up. Since we start from a small phase-shift value in the carrier, the retardation of maximal wave focusing of the waves would require a long fetch to observe interesting post-focusing dynamics (recurrence, triangular patterns etc.). This is also the case for the KM breather. We chose the case of amplitude focusing factor of 3.3 and as the previous breather cases, the two type of wave envelope dynamics are shown in Fig. \ref{fig4}. 
We annotate that we employ the same colobar scale in all Figures to make significant wave focusing more distinct. Overall, all cases show a reasonable good agreement with the weakly nonlinear NLSE framework approximated from the Euler equations to third-order in steepness. That said, a typical wave asymmetry is clearly noticeable in the experimental data. The latter can be explained and modelled by the modified NLSE accounting for fourth-order effects in steepness in form of higher-order dispersion and mean flow \cite{trulsen2001spatial,shemer2013peregrine,Slunyaevetal2013}. Note that when injecting the PSP an adjustment of the linear wave maker's frequency-related transfer function is required to ensure the regularity of the wave train. 

In summary, we have reported an experimental study confirming a proof of concept that rogue waves can appear from a localized PSP in a regular wave train. Such PSP can be for instance constructed from breather solutions of the nonlinear Schrödinger equation and then, seeded in a condensate. Our wave tank measurements underline that nonlinear focusing can be exhibited by PSP confirming yet another focusing mechanism beyond the wave modulation or superposition principle. We anticipate future experimental studies exploring such focusing dynamics in various wave systems governed by dispersion and nonlinearity \cite{kourakis2013semiclassical,luo2020creation,frisquet2013collision} and also the development of theoretical techniques to predict such dynamics within the NLSE framework and beyond \cite{suret2020nonlinear,mullyadzhanov2021solitons}. 


\bibliography{Refs}

\end{document}